# Modelling the emergence of
# spatial patterns of economic activity

**Jung-Hun Yang, Dick Ettema, Koen Frenken**

Urban and Regional Research Centre, Faculty of Geosciences, Utrecht University
P.O.Box 80115, Utrecht, 3508 TC, the Netherlands
e-mail: j.yang@geo.uu.nl,  d.ettema@geo.uu.nl,  k.frenken@geo.uu.nl

## 1. Introduction

The spatial pattern of economic activities is an important determinant of urban development. Locations of firms influence where workers will live, where consumers will buy products and where other firms are located. The locations of firms also impact on transportation flows, since they are important attractors and producers of both personal and freight traffic. Finally, the spatial pattern of firms obviously has a profound impact on the economic viability and conditions for economic growth in a region. Through the decades, therefore, researchers have developed models that describe and predict how spatial patterns of economic activity emerge. Without intending to exhaustively review all approaches taken, we will here review some modelling approaches that are relevant to our study. In particular, we will review micro-simulation and agent based approaches that take the individual firm as the unit of analysis.

A first type of models (UrbanSim, SimFirms, ILUMASS) describes the evolution of spatial economic systems as a stochastic process, in which events such as firm growth, firm relocation, spin offs and take place with a probability that is predominantly a function of firm characteristics. In UrbanSim, economic activity is represented in terms of individual jobs, which are taken from an independent economic forecasting model, and are exogenous to the model. The jobs are treated as independent entities (i.e. not organised in firms), which are distributed across grid cells. ILUMASS (Moeckel, 2005) applies a more elaborate economic component. In particular, it uses a synthetic database of firms, which may take decisions regarding relocation, growth and closure. In addition, new firms may emerge at a particular birth rate, which is specific per sector and dependent on general economic growth rates. One of the most elaborate micro-simulation models of firms' developed to date is SIMFIRMS (Van Wissen, 2000). This model distinguishes the same events as ILUMASS (birth, growth, (re-) location, closure) but uses more sophisticated behavioural rules, accounting for such factors as market stress, spin offs of existing firms, age effects and spatial inertia in the case of relocation. Market stress is related to the concept of carrying capacity, which, analogous to the ecological concept, indicates the maximum number of firms that an urban system can contain. Carrying capacity is operationalised as the difference between market supply and market capacity, which is based on aggregate input-output models. Thus, the measure is the outcome of aggregate conceptualisations, rather than on firms' perception of demand and supply. In general the micro-simulation approaches are especially insightful to study demographic processes. For instance, they suffice to describe what the distribution across sectors in a region will be given some initial setting and given birth rates, spin-off probabilities etc. An element that is much less developed in these models is the role of spatial proximity. The fact that firms cluster in order to achieve agglomeration advantages is not well represented. Structural changes in spatial economics structures (e.g. the emergence of new economic centres due to changes in industries) are not well represented.

A second type of models focuses on the emergence of hierarchies of concentrations (of firms or population) as a result of simple reproduction and migration rules. Simon (1955) shows that by assuming fixed reproduction rates and relocation probabilities, and assuming that larger concentrations attract more migrants than lower concentrations, a hierarchy of concentrations emerges that follows a power law distribution. Remarkably, such power law distributions match existing hierarchies in economic concentration (Frenken et al., 2007) and population concentrations (Pumain, 2006) very well. Although apparently these simple reproduction and migration rules touch upon general principles of spatial organisation, the theoretical underpinning of the models is somewhat cumbersome (Krugman, 1996). In their most basic form, models as suggested by Simon are non-spatial. That is to say, the relative position of a concentration (e.g. a city or a commercial area) to other concentrations does not matter, since locational preferences of migrants only depend on the size of the concentration and not on its surroundings. As a result, a big city on an isolated place would be equally attractive as an equally big city surrounded by other cities. This assumption is problematic since it ignores the impact of proximity. For instance, studies in evolutionary economics (Boschma et al., 2002) suggest that proximity to other firm maters for their productivity and innovative capacity, and that this proximity exceeds the purely local scale. In particular, regions play an important role in processes of economic innovation, where the size of a region differs between types of industries. Thus, although correctly reproducing the rank size distribution of existing economic and population concentrations, the Simon model falls short in describing the emergence of clusters of economic development on a regional level.

From the above, we conclude that existing models of spatial economic development have some important limitations. Most importantly, the role of spatial proximity to other firms is not well represented in the models. This proximity includes both the availability of other firms and the distance to these firms. Given this shortcoming, the objective of this paper is to propose a model of spatial economic development that is capable of representing the impact of spatial proximity on emerging spatial patterns. To this end, a theoretical framework is developed in which market potential, agglomeration benefits and congestion affect locational decisions on different spatial scales. The model of location behaviour is embedded in a demographic model of firm growth and spin-off processes. The model is tested in a stylised spatial setting, to illustrate how different parameter settings lead to different spatial configurations.

The paper is organised as follows. Section 2 outlines a model that describes the behaviour divisions and the utility of spatial proximity. Section 3 describes the application of the model in a series of simulations. Section 4 analyses the impacts of spatial proximity, weighting effects and relocation probabilities on the emerging patterns of economic activity.

## 2. Model description

In line with the models reviewed above, our model describes the spatial behaviour of firms dynamically over a number of time steps. However, as firms may consist of multiple establishments and divisions, that may take individual locational decisions, we take the division as the unit of analysis. We define divisions as coherent working units with a minimum and maximum size, dependent on the type of firm. The behaviours described by the model are growth, spin-off and relocation. With respect to internal growth, we assume that divisions in a certain sector grow uniformly with a fixed amount per year. In reality, growth rates will differ between firms due to factors such as quality of management, position in a network of firms and geographical position. Although we recognise the existence of such heterogeneity, we will not include it in this study. In particular, we assume that division size in year *t+1* equals:

$$Division(InternalGrowth) : Div_{t+1}^{size} = Div_t^{size} + 1 \qquad (\text{ex.1})$$

where $Div_{t+1}^{size}$ is the size of a division in a year *t+1*. We further assume that divisions have a maximum size and that growth beyond this maximum results in less effective functioning of divisions, e.g. through increasing overheads. Hence we assume that if the maximum size is reached the division will split, resulting in a new division (spin-off). To reflect developments in product and sector lifecycles a spin-off does not necessarily result in a division of the same type as the parent division. For instance, a spin-off of an industrial division may be a division in services or high-tech. This reflects ongoing shifts in economies from traditional industries to high-tech and from manufacturing to services. In the models tested in this study we will assume the existence of a traditional and an innovative industry, in which all spin-offs (both from traditional and new industries) are innovative industries. The rule for occurrence of spin-offs is:

$$Division(Spinoff)$$
$$= \begin{cases} SpinoffDiv_{t+1}^{type} = old \,\&\, Div_{t+1}^{size} = 0 \_ if((Div_t^{type} = old)\,\&\,(Div_t^{size} > \delta_{old})) \\ SpinoffDiv_{t+1}^{type} = new \,\&\, Div_{t+1}^{size} = 0 \_ if((Div_t^{type} = new)\,\&\,(Div_t^{size} > \delta_{new})) \\ SpinoffDiv_{t+1}^{type} = new \_ if((Div_t^{type} = old)\,\&\,(Div_t^{size} = \delta_{old})\,\&\,(Div_t \in random(\phi))) \end{cases} \qquad (\text{ex.2})$$

where $Div_{t+1}^{type}$ is the type of division in time step $t+1$ and $SpinoffDiv_{t+1}^{type}$ is the type of division which is split. $\delta_{old}$ and $\delta_{new}$ are constants indicating the maximum size of divisions of old and new type respectively. The first and second line of *expression 2* imply that if the maximum size is reached the division is split as the same type and size and is initialised with size zero. If an old type division reaches its maximum size with probability $\phi$, the spin-off is the new type.

Apart from such demographic processes, the model describes firms' relocation behaviour. Relocation of firms may take place for many reasons, which are usually concerned with internal processes, such as growth or suitability of the building. In such cases, the relocation is likely to take place within the same municipality or region, without structurally changing the spatial structure of the economy. In this study, however, we are particularly interested in the more strategic relocation, in which divisions seek to improve their access to markets and resources by moving to another geographic location. In this respect, we assume that each division has a certain probability to evaluate its current geographical position against alternative positions to test whether relocation results in an improvement of its conditions. Two options are distinguished. First, a division may investigate relocation to an existing city (defined as an existing concentration of firms) or to a new place without a current concentration of firms. We assume that the probabilities of not exploring relocation, investigating relocation to an existing and to a new city are 0.9, 0.09 and 0.01 respectively. It is recognised that different probabilities of relocation may result in different spatial patterns and different development speeds. This relation can be described as:

$$W_{t+1} = W_t + \kappa \cdot W_t \tag{ex.3}$$

$$= W_t + (1-\pi) \cdot \kappa \cdot W_t + \pi \cdot \kappa \cdot W_t \tag{ex.4}$$

$$= (1-\gamma) \cdot (W_t + (1-\pi) \cdot \kappa \cdot W_t) + \gamma \cdot (W_t + (1-\pi) \cdot \kappa \cdot W_t) + \pi \cdot \kappa \cdot W_t \tag{ex.5}$$

$$= \lambda_1 \cdot (W_t + (1-\pi) \cdot \kappa \cdot W_t) + \lambda_2 \cdot (W_t + (1-\pi) \cdot \kappa \cdot W_t) + \lambda_3 \cdot \kappa \cdot W_t \tag{ex.6}$$

where $W_t$ is the total number of divisions in a year $t$ and $\kappa$ is the growth rate of the industry. The second term of *expression 4*, which plays an important role in the persistence and self-reinforcement of clusters, is the number of division that is attached to an existing city by $1-\pi$ and the third term is the number of divisions moving to a new city (vacant area) based on the rate $\pi$. $\gamma$ is the rate of moving to another city. The first term of *expression 5* is the number of divisions not moving to another city and the second term is the number moving to another city. The third term is the number of divisions moving to new city. $\lambda_1$, $\lambda_2$ and $\lambda_3$ are the probabilities of not exploring relocation, moving to an existing and to a new city respectively. $\lambda_1$ and $\lambda_2$ can be matched with the $1-\gamma$ and $\gamma$. $\lambda_3$ is also equal to $\pi$ of *expression* 5. However, since the focus of our study is on the role of spatial proximity in the emergence of spatial patterns, we use the above values, which proved to work well in other studies. The rule of relocation can be therefore defined as:

$$Division(relocation) = \begin{cases} Div_{t+1}^{move} \leftarrow Don'tMove\_if(Div_t \in random(\lambda_1)) \\ Div_{t+1}^{move} \leftarrow \max_{m=1}(U_m)\_if(Div_t \in random(\lambda_2)) \\ Div_{t+1}^{move} \leftarrow \max_{n=1}(U_n)\_if(Div_t \in random(\lambda_3)) \end{cases} \quad (ex.7)$$

where $U_m$ is the utility of city $m$ and $U_n$ is the utility of vacant area $n$.

Evaluation of alternative locations and relocation take place as follows. A firm will evaluate all locations to find the location with the highest utility. If the utility is higher than the utility of the current location, the division will move to this new location, otherwise, it will stay in its current place.

The central issue when discussing the impact of spatial proximity is how utility is defined. In a non-spatial model, utility of each location would be equal, suggesting a random spatial process. The spatial sensitivity of the model is improved if the locational preference depends on the size (number of divisions) in the destination. In this case utility is defined as:

$$U_i = N_i \quad (ex.8)$$

where $U_i$ is the utility of area $i$ and $N_i$ is the number of division in an area $i$. In essence, this is the model proposed by Simon, which leads to the well known power law distribution of concentrations.

As noted before, the Simon model is local in terms of its utility function, since it only accounts for firms in a certain location, and not in the surroundings are taken into account. In this respect, this study aims at proposing and testing utility formulations that not only take into account locational characteristics, but also characteristics of the surroundings, such as the proximity to other firms. Looking at locational characteristics of firms, the literature suggests various factors relating to proximity of firms that clearly exceed the purely local level.

A first factor concerns market potential. Firms make profits from selling products of services to other firms or to individuals. The shorter the travel distance to these clients, the lower the costs and the higher the profit. In addition, the more clients can be reached within acceptable travel distance from a location, the larger the market potential and the more attractive the location is to settle. In this respect, the sensitivity to distance is the factor determining the spatial configuration. For common goods, such as groceries, willingness to travel is low. For more specialised goods/services, the willingness to travel and the market area will be larger. Such differences in distance decay will have a large impact on the emerging spatial patterns of economic activity. To operationalise this factor we assume that firms from different sectors buy each others products and that firms also serve as a proxy for the number of consumers that are wiling to buy goods or services. Hence, market potential (MP) can be defined as:

$$MP_i = \sum_{j=1} N_j e^{-\alpha_1 d_{ij}} \qquad (ex.9)$$

where $MP_i$ is the market potential in area $i$ and $N_j$ is population (number of divisions) in city $j$. $\alpha_1$ is a parameter for controlling the distance decay and $d_{ij}$ is distance between area $i$ and city $j$. A second factor related to spatial proximity is agglomeration advantages. Many studies suggest that firms benefit from proximity to similar firms. One reason is that they may profit from shared facilities and suppliers. In addition, some firms may be better able to attract clients and employees jointly than individually. Another important issue is that firms form networks in which knowledge is exchanged, projects are carried out and market information is exchanged, in order to achieve competitive advantages. Such agglomeration advantages suggest that firms will prefer to locate near other firms from the same sector. In equation, agglomeration effects are expressed as:

$$AP_i^{type} = \sum_{j=1} N_j^{type} e^{-\alpha_2 d_{ij}} \qquad (\text{ex.10})$$

where $AP_i^{type}$ is the agglomeration potential of division of specific type in area $i$ and $N_j^{type}$ is a population of divisions of a specific type in city $j$. $\alpha_2$ is a parameter for controlling the distance effect and $d_{ij}$ is distance between area $i$ and city $j$. It is recognised that agglomeration advantages for different sectors may differ in importance, e.g. due to the relative importance of knowledge and innovation in a sector. Also the scale of agglomeration advantages may differ, due to the type of interaction. E.g. having similar consumers asks for immediate physical proximity, whereas exchange of knowledge via personal meetings allows a longer travel time.

Finally, having noted the advantages of being close to other firms and clients, we note that there will also be disadvantages. Increasing density leads to congestion of infrastructure and facilities, but also to higher prices and increasing competition for employees and other resources. Note that congestion is not sector specific, in the sense that firms suffer from congestion caused by all other firms. In equations:

$$CP_i = \sum_{j=1} N_j e^{-\alpha_3 d_{ij}} \qquad (\text{ex.11})$$

where $CP_i$ is the congestion effect in area $i$ and $N_j$ is population in city $j$. $\alpha_3$ is a parameter for controlling distance effect and $d_{ij}$ is distance between area $i$ and city $j$. Again, we note that sensitivity or congestion may differ between firm types, due to their need for space and infrastructure and the required qualifications of their employees. However, also the advantage of agglomeration will be weighted off against the disadvantage of congestion. As a result, the Simon utility of *expression 8* can be transformed as:

$$U_i = \beta_1 MP_i + \beta_2 AP_i^{type} + \beta_3 CP_i \tag{ex.12}$$

$$= \beta_1 \sum_{j=1} N_j e^{-\alpha_1 d_{ij}} + \beta_2 \sum_{j=1} N_j^{type} e^{-\alpha_2 d_{ij}} + \beta_3 \sum_{j=1} N_j e^{-\alpha_3 d_{ij}} \tag{ex.13}$$

where $\beta$ is a coordinating parameter and $MP_i$ is the market potential in an area $i$. $AP_i^{type}$ is the agglomeration potential of type $t$ and $CP_i$ is the congestion potential in the area $i$. The relocation probability can then be defined as:

$$P_i = \frac{\exp(U_i)}{\sum_{j=1} \exp(U_j)} = \frac{\exp(\beta_1 MP_i + \beta_2 AP_i^{type} + \beta_3 CP_i)}{\sum_{j=1} \exp(\beta_1 MP_j + \beta_2 AP_i^{type} + \beta_3 CP_j)} \tag{ex.14}$$

where $P_i$ is the probability of area $i$ for relocation. This function is applied both for migration to existing cities and for migration to new cities.

To summarise, our model assumes that apart from locational and building specific characteristics, firms' locational preferences are guided by various variables that express the proximity to other firms, of similar and other sectors. In particular, we assume that market potential, agglomeration advantages and congestion effects, as defined in the above influence more strategic decisions about where firms are located. At the same time, we assume that firms differ with respect to the importance of these effects and the spatial scale at which they play. We hypothesize that the preferences of firms with respect to proximity will determine the spatial configuration of economic activities. For instance, agglomeration advantages on a small scale will lead to multiple centres of economic activity, whereas agglomeration advantages on a larger scale may lead to a single centre. In the remainder of this paper we will test to what extent differences in the spatial scale will lead to different spatial configurations.

## 3. Study design

The objective of this study is to test to what extent differences in firms' preferences with respect to spatial proximity lead to different spatial patterns of economic activities.

In addition, we want to find out to what extent the impact of firms' preferences is affected by factors such as growth rates per sector and the flexibility of relocation. Although we recognise that many factors other than discussed before (such as the availability of facilities, path dependency etc.) impact on firms' location choice, we will use a stylised setting in which we will test some fundamental relationships between individual preferences of firms on the one hand, and aggregate spatial patterns on the other hand. In particular, we assume that firms operate in a landscape that is homogeneous in terms of travel speeds and quality of locations, and only varies in terms of the presence of other firms. The landscape consists of a square of 50x50 cells. Initially, at t=0, the landscape is filled with 2500 divisions, which in each time step will grow and with some probability relocate. The likelihood and effectuation of these events is determined by the equations described in the above. To test the impact of different preferences of spatial proximity, the model will be run with different parameters during 210 Time steps, after which the resulting pattern is analysed. This analysis will include three elements.

First, the resulting patterns will be interpreted visually in terms of the number and size of emerging clusters of economic activity. Second, the distribution of rank sizes will be plotted, to see whether the resulting patterns follow the power law distribution typical for urban and economic distributions (Simon, 1955; Pumain, 2006). Third, the degree of clustering is expressed using the formula:

$$K^{extension}(d) = \frac{\sum_{i=1} N_i \cdot no[S \in C(s_i, d)]}{N} \quad \text{(ex.15)}$$

$$L(d) = \sqrt{\frac{K^{extension}(d)}{\pi}} - d \quad \text{(ex.16)}$$

where $K^{extension}$ is the cluster density and $N$ is the total number of divisions. $C(s_i, d)$ is a circle with distance $d$ from $s_i$. This index has a higher value if more divisions are closer to one another. The index is calculated with the distance 10 for the purpose of our study.

Starting point of the analyses is a base specification of the model, with parameters specified as in Table 1. Relative to this base model, the following analyses are carried out. First, starting from a Simon-type model without locational preferences (beta are zero), various spatial factors are added stepwise, to see how this changes the resulting pattern. Second, the impact of different spatial factors will be varied by changing the beta parameter, in order to find out how this relative impact affects the spatial pattern of economic activity. Finally, the base model will be run with varying values for the parameters lambda, to see how that influences the resulting spatial pattern.

## 4. Simulation results

As shown in figure 1, the initial state of simulation is that the division of old type is equally distributed across all regions which consists of 2500 cells (50 by 50). A division in the old industry may grow in each time step (*ex.1*) leading to a spin-off (*ex.2*) or relocation (*ex.7*). The utility and probability for migration also follows the *expressions 13* and *14* respectively. The figure 1 shows that the spatial pattern of old industry changed from an evenly distributed pattern into a clustered spatial pattern. Concerning the emergence of a new industry, we assume a growth rate five times higher than that of the old industry. The different parameters for the old and new industry clearly affect their pattern of evolution. The new industry emerges both in a new agglomeration and in the existing agglomeration. This can be understood from the fact that the new industry profits from agglomeration economies of co-location (which explains the emergence of the new agglomeration) as well as from proximity to demand (explaining the growth of the new industry in the existing agglomeration). The resulting spatial pattern after the new industry has emerged, has become more "Zipf-like" in the sense that we can witness cities of different sizes with the frequency of particular size decreasing with increasing size. Our model thus underlines the need to understand the spatial structure of an economy as a historical process of structural change leading to a progressive diversification of the economy. Figure 2 shows the cluster degree of each model run. Models 2 to 4 of red colour have a higher value than models 5 to 7 of green colour, though differences are small. The model 1 has the lowest value because any spatial interaction structures are absent. In addition, the total population of division represents the exponential growth.

**Table 1. Parameter for Model 1~7**

| Parameter | | Type | Model 1 | 2 | 3 | 4 | 5 | 6 | 7 |
|---|---|---|---|---|---|---|---|---|---|
| | | | Empty | Only MP | MP+ AP | MP+ AP+ CP | Larger MP | Larger AP | Larger MP+ AP |
| MP | $\alpha_1$ | Old | 0.5 | 0.5 | 0.5 | 0.5 | 0.5 | 0.5 | 0.5 |
| | | New | 0.4 | 0.4 | 0.4 | 0.4 | 0.2 | 0.4 | 0.2 |
| AP | $\alpha_2$ | Old | 0.5 | 0.5 | 0.5 | 0.5 | 0.5 | 0.5 | 0.5 |
| | | New | 0.4 | 0.4 | 0.4 | 0.4 | 0.4 | 0.2 | 0.2 |
| CP | $\alpha_3$ | Old | 0.5 | 0.5 | 0.5 | 0.5 | 0.5 | 0.5 | 0.5 |
| | | New | 0.4 | 0.4 | 0.4 | 0.4 | 0.4 | 0.4 | 0.4 |
| MP ct | $\beta_1$ | Old | 0 | 1 | 1 | 1 | 1 | 1 | 1 |
| | | New | 0 | 1 | 1 | 1 | 1 | 1 | 1 |
| AP ct | $\beta_2$ | Old | 0 | 0 | 0.5 | 0.5 | 0.5 | 0.5 | 0.5 |
| | | New | 0 | 0 | 0.5 | 0.5 | 0.5 | 0.5 | 0.5 |
| CP effect | $\beta_3$ | Old | 0 | 0 | 0 | −1 | −1 | −1 | −1 |
| | | New | 0 | 0 | 0 | −1 | −1 | −1 | −1 |
| Ma Growth | $\delta_{old}$ | Old | 50 | 50 | 50 | 50 | 50 | 50 | 50 |
| | $\delta_{new}$ | New | 10 | 10 | 10 | 10 | 10 | 10 | 10 |
| Migration | $\lambda_2$ | ot y | 19 | 19 | 19 | 19 | 19 | 19 | 19 |
| | $\lambda_3$ | new city | 0.3 | 0.3 | 0.3 | 0.3 | 0.3 | 0.3 | 0.3 |
| Time Step | - | | 210 | 210 | 210 | 210 | 210 | 210 | 210 |

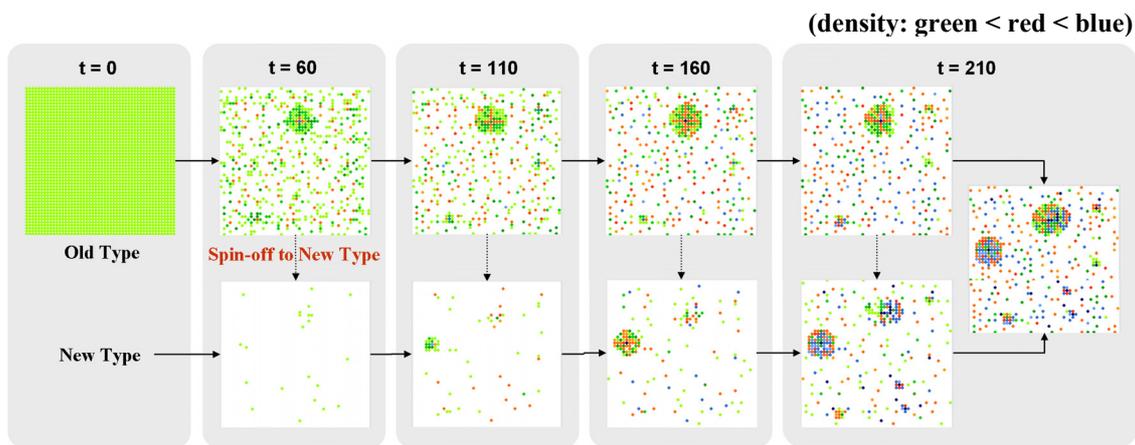

**Fig. 1. Time series for spatial pattern of old and new type division
(based on $\lambda_2 = 7$, $\lambda_3 = 0.2$ and Model 4)**

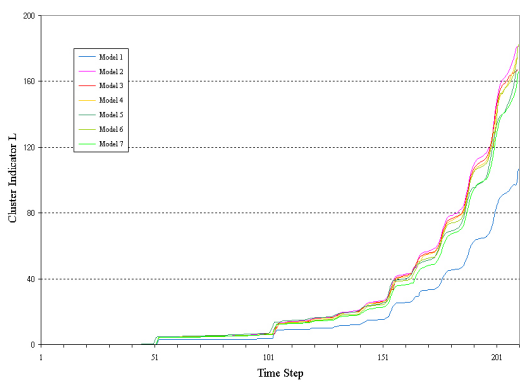

**Fig. 2. Time series for cluster indicator L of Model 1~7**

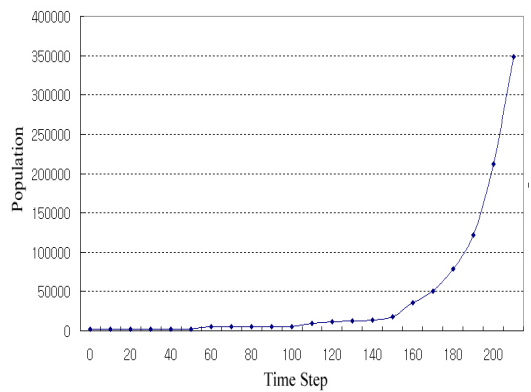

**Fig. 3. Time series for population of division**

### 4.1. The impact of spatial proximity

The first model (*model 1*) that is tested is only based on growth and spin-off processes, lacking spatial preferences. This model is rather similar to the Simon model, except for the fact that Simon's model assumes that cells with more divisions are more likely to attract newcomers, whereas in our model all cells have equal probability. As seen in figure 4, this model results in a pattern without centres, with divisions scattered out over space and filling all cells.

The second model (*model 2*) includes the proximity to both old and new industry firms, representing market potential. The second picture of figure 4 suggests that adding market power results in a more clustered configuration, with one large centre. In time step 210, two smaller subcentres have emerged, which may in time develop to new centres. The rank size distribution clearly shows the power law distribution in figure 5.

The third model (*model 3*) includes market potential and agglomeration effects, where agglomeration effects only relate to proximity to similar firm types. The simulation suggests that this model also leads to clustering of divisions, with one large centre. However, agglomeration may, since it is only focussed on similar firm types, more easily result in local clusters, such as the cluster in the lower left corner. As a consequence, the cluster indicator L is lower than in the case where only market potential plays a role. Again, the rank size distribution represents the zipf distribution more than a model 2.

Finally, (*model 4*) adding the impact of congestion to the model results in a pattern with one centre like the fourth picture of figure 4. Remarkably, this pattern is less fragmented than the model with market potential and agglomeration effects, although the congestion is supposed to lead to more dispersed locations.

Overall, we observe that centripetal forces such as market potential and agglomeration lead to a higher degree of clustering of economic activity in our model. The effect of congestion, however, is limited.

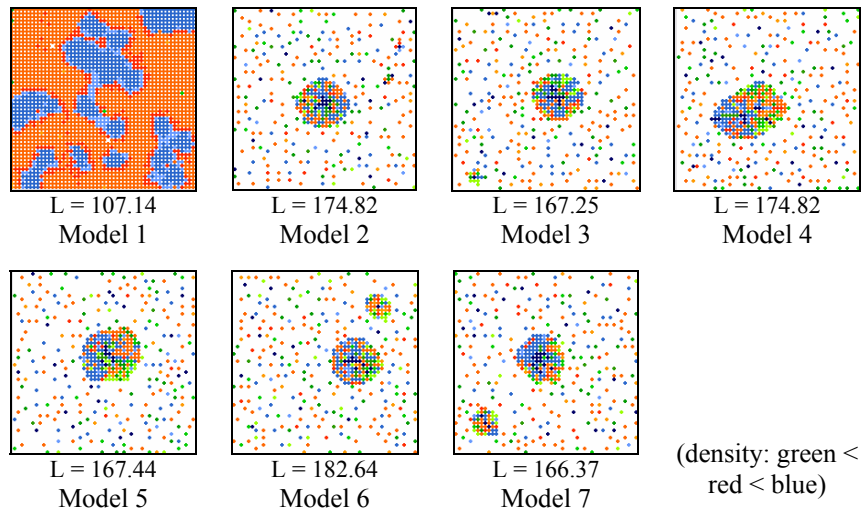

**Fig. 4. Spatial Pattern: The impact of spatial proximity and weighting effects**

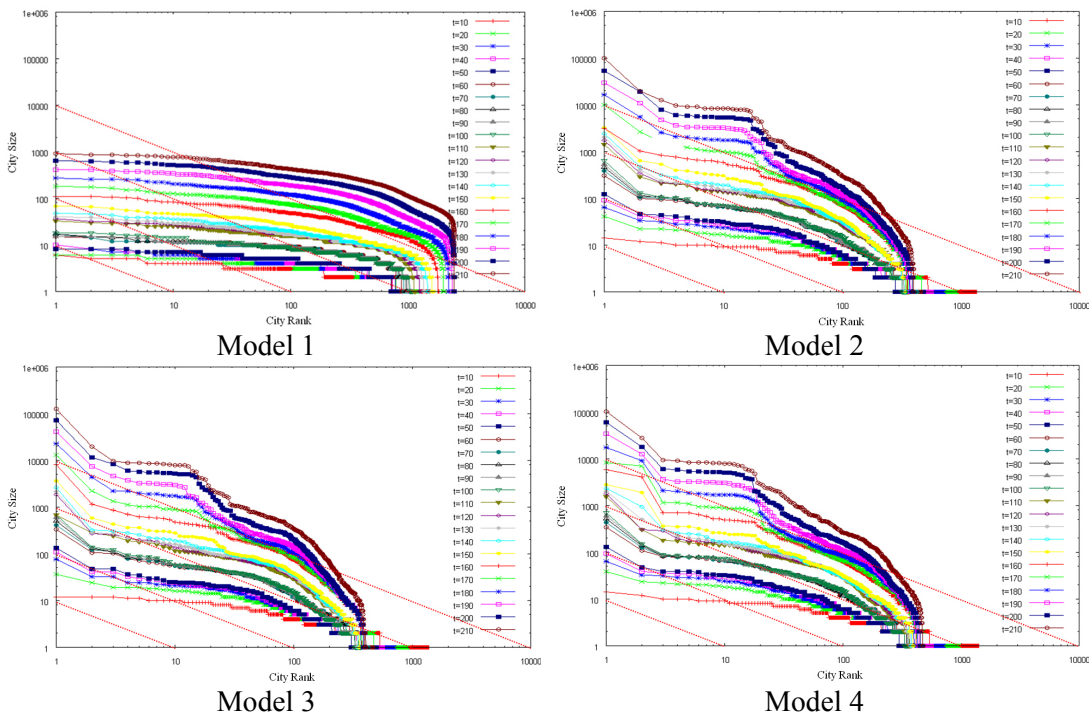

**Fig. 5. Rank size distribution**

*4.2. Weighting effects*

In this section various models in which the relative impact of market potential and agglomeration are varied, are discussed. As seen in figure 4, Model 5 has a lower $\alpha_1$ for the market potential of new industries, implying that products are delivered in a larger area (e.g. because transport costs are lower). This model results in a spatial pattern with one large centre, which is visually hard to distinguish from the base model. The cluster index, however, suggests that the degree of clustering is lower as compared to the base model. This would be logical, given that the market area is larger, reducing the need to be in the immediate proximity of clients.

Model 6 increases the spatial scale of agglomeration effects through a lower $\alpha_2$ for the agglomeration potential of new industries. This results in a pattern with one larger and one smaller centre. Apparently, the larger spatial reach increases the attractiveness of less densely 'populated' areas, increasing also the probability of subcentres emerging. Yet, the cluster index has a high value, suggesting that within and around the clusters accessibility to other firms is high. This is also the result of the relative closeness of the clusters.

Model 7, in which both market potential and agglomeration potential have a larger spatial reach, clustering becomes less, as expected. The larger spatial reach facilitates the emergence of two clusters that are more equal in size than in the other models.

To conclude, the spatial reach of the three identified effects has an impact on the emerging pattern. In particular, it seems more likely that multiple clusters emerge, since market and agglomeration advantages are available in a larger area. The impact this has on the cluster index varies, depending on the locations where clusters emerge and also depending on the scale of the cluster index.

*4.3. The impact of relocation probabilities*

To test the impact of relocation probabilities, the variables $\lambda_2$ and $\lambda_3$ were varied to represent a higher relocation probability in general (mostly determined by $\lambda_2$) and a higher probability of moving to a new city ($\lambda_3$). For each combination, the emerging spatial pattern as well as the cluster index are displayed (figure 7). Visual inspection of the emerging patterns suggests that especially $\lambda_3$ has a significant impact on the outcomes. In particular, a higher probability of moving to a new city results in a setting with one large centre without subcentres, whereas a lower value of $\lambda_3$ leads to more subcentres. Apparently, constraining the opportunity to settle down in a new city increases the probability that subcentres emerge in existing cities. In most cases, increasing the probability of moving to a new city ($\lambda_3$) results in a higher degree of clustering (figure 6 and table 2), as a result of the more centralised configuration. It is noted however, that this effect is less obvious in case of a lower probability of moving to an existing city ($\lambda_2$).

**Table 2. The average of indicator L by 4 times simulation based on Model 4**

| $\lambda_3$ | 0.2 % | 0.3 % | 0.4 % | 0.5 % |
|---|---|---|---|---|
| 7 % | 148.44 | 149.90 | 188.54 | 228.66 |
|  | 148.91 | 152.33 | 176.21 | 245.22 |
|  | 151.93 | 162.70 | 202.97 | 219.69 |
|  | 153.76 | 165.46 | 209.16 | 217.42 |
| average | 150.76 | 157.60 | 194.22 | 227.75 |
| 11 % | 146.47 | 155.31 | 200.71 | 210.01 |
|  | 162.06 | 169.77 | 174.95 | 201.90 |
|  | 157.11 | 156.50 | 212.44 | 218.96 |
|  | 149.83 | 167.05 | 190.31 | 193.74 |
| average | 153.87 | 162.15 | 194.60 | 206.15 |
| 15 % | 121.67 | 174.38 | 201.53 | 226.93 |
|  | 136.33 | 190.49 | 209.07 | 163.68 |
|  | 164.42 | 187.67 | 181.41 | 225.91 |
|  | 153.07 | 138.46 | 202.36 | 205.03 |
| average | 143.87 | 172.75 | 198.59 | 205.39 |
| 19 % | 164.92 | 160.58 | 214.86 | 255.57 |
|  | 153.78 | 191.16 | 186.33 | 205.20 |
|  | 144.99 | 156.58 | 206.44 | 171.43 |
|  | 128.67 | 170.60 | 174.24 | 208.32 |
| average | 148.09 | 169.73 | 195.47 | 210.13 |

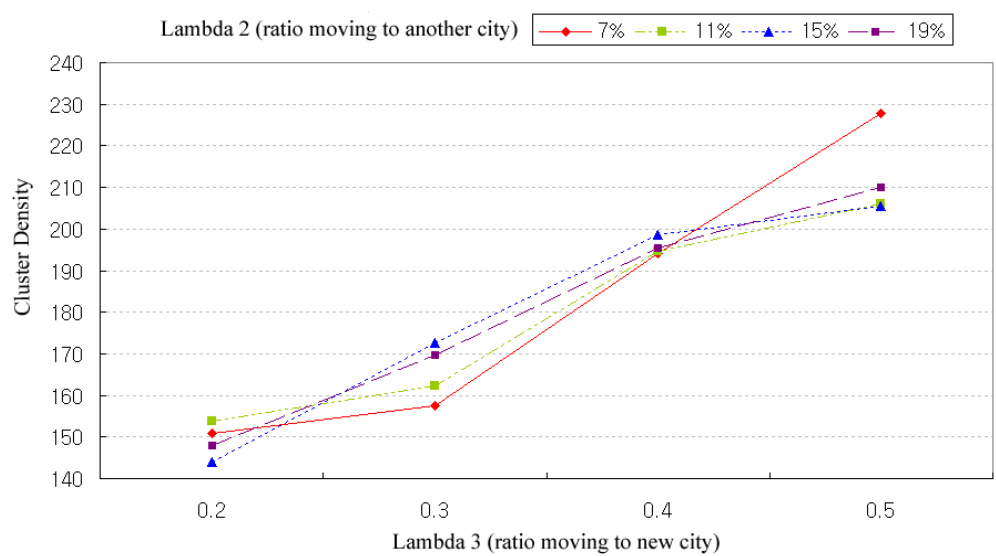

**Fig. 6. Indicator L by $\lambda_2$ and $\lambda_3$ based on Model 4 (ref. Table 2)**

## 5. Conclusion and discussion

In this paper we have demonstrated in a stylised setting, how differences in preferences with respect to spatial proximity lead to different spatial patterns of economic activity.

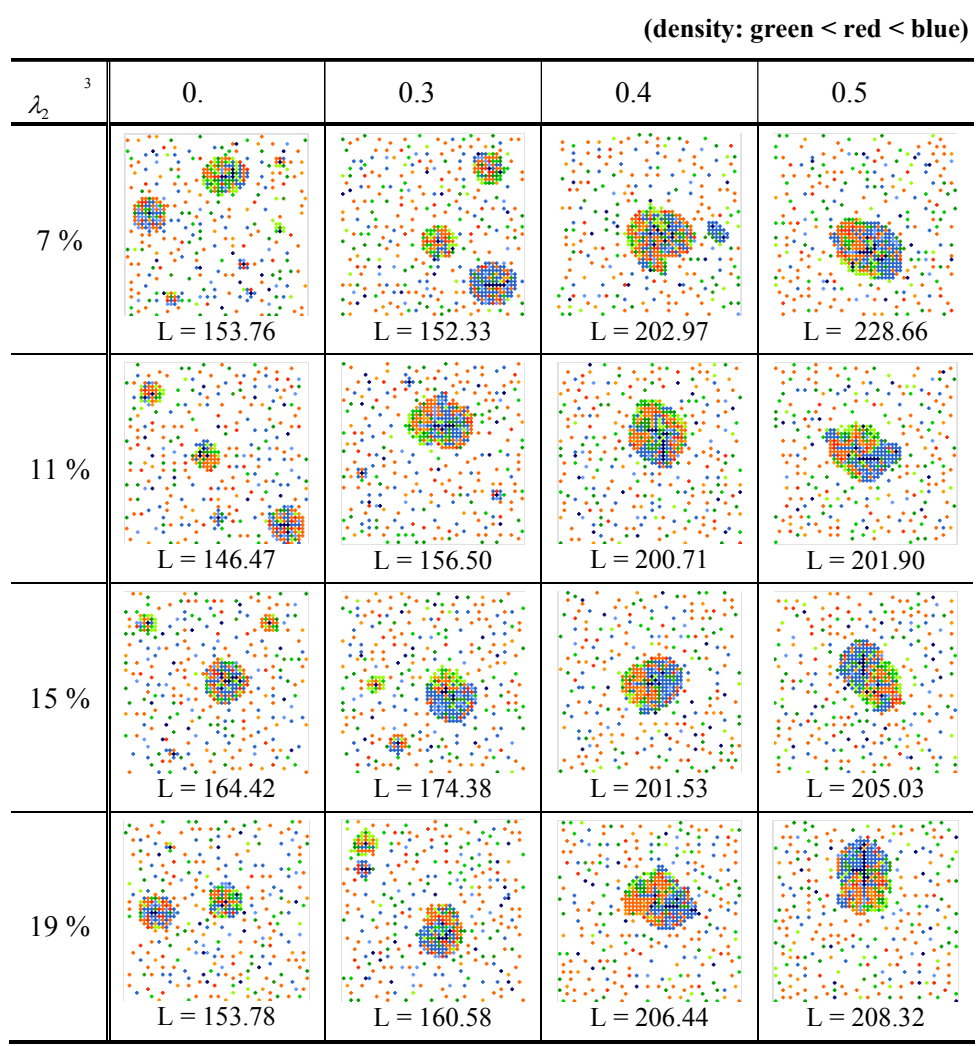

Fig. 7. Spatial Pattern: The impact of relocation probabilities (based on Model 4)

In this respect, a preference to achieve a high market potential and to profit from agglomeration advantages results in more centralised settings. However, the spatial scale of the market and agglomeration effects matters. In particular, if agglomeration advantages stretch out over a longer distance, more subcentres emerge. Somewhat surprisingly, congestion seems to have a minor impact on the emerging patterns. Although the simulation outcomes are intuitively plausible, they also articulate the need for validation of the behavioural decision rules. If outcomes are determined by the presence (and potentially strength, although not tested in this paper) and spatial reach of market potential, agglomeration and congestion effects, it is important to investigate how firms of different types valuate these factors in their location choice behaviour. In particular, it is important how the valuation of these factors varies with firm characteristics such as type of activities, size, history and the position in economic networks. Such information would be necessary to apply the above approach in a more realistic setting as a policy support tool. A second conclusion that can be drawn from the simulations is that relocation probability to existing and new cities impacts on the emerging patterns. This finding is highly policy relevant, since it suggests that the availability of locations where firms/divisions can move has a significant impact on spatial patterns of economic activity. If this is confirmed by validation studies, it would suggest that spatial planning is a tool that can directly impact on the economic structure of regions and will influence firms' performance and thereby regional economic development.

Although this study provides first insights into the emergence of spatial patterns of economic activity, it is obvious that much more research is needed to develop this approach into a tool that can be readily used for policy analysis. This research should address the following issues. First, the behavioural rules applied in this test of concept need to be verified and refined. In particular, multivariate analyses are needed that relate firm characteristics to the degree and spatial reach of proximity preferences. This will require dedicated data to be collected from individual firms. Second, it should be recognised that firms do not operate in isolation, but interact with households and individuals (as clients and employees), institutions (such as government agencies, universities, schools etc.) and react to the physical environment (landscape, quality of residential environment, pollution and noise). A proper model for policy evaluation should include a representation of how proximity concerns are traded off against these other factors.